\documentclass[conference]{IEEEtran}

\usepackage{amsmath}
\usepackage{url}
\usepackage{graphicx}
\usepackage{color}
\usepackage{placeins}
\usepackage{float}
\usepackage{tabularx,colortbl}
\usepackage{xfrac}
\usepackage[autostyle, english = american]{csquotes}
\usepackage[table,xcdraw]{xcolor}    
\usepackage{booktabs}
\usepackage{multirow}
\usepackage{microtype}  
\newcommand\blfootnote[1]{%
  \begingroup
  \renewcommand\thefootnote{}\footnote{#1}%
  \addtocounter{footnote}{-1}%
  \endgroup
}

\begin{document}

\title{5G NR Jamming, Spoofing, and Sniffing:\\Threat Assessment and Mitigation}

\author{\IEEEauthorblockN{Marc Lichtman\textsuperscript{1}}
\IEEEauthorblockA{Vencore Labs\\
Basking Ridge, NJ, USA\\
Email: mlichtman@vencorelabs.com}
\and
\IEEEauthorblockN{Raghunandan Rao, Vuk Marojevic, Jeffrey Reed}
\IEEEauthorblockA{Wireless@VT, Virginia Tech\\
Blacksburg, VA, USA\\
Email: \{raghumr,maroje,reedjh\}@vt.edu}
\and
\IEEEauthorblockN{Roger Piqueras Jover}
\IEEEauthorblockA{Bloomberg LP\\
New York, NY, USA\\
Email: rpiquerasjov@bloomberg.net}}

\maketitle

\begin{abstract}
In December 2017, the Third Generation Partnership Project (3GPP) released the first set of specifications for 5G New Radio (NR), which is currently the most widely accepted 5G cellular standard.  5G NR is expected to replace LTE and previous generations of cellular technology over the next several years, providing higher throughput, lower latency, and a host of new features.  Similar to LTE, the 5G NR physical layer consists of several physical channels and signals, most of which are vital to the operation of the network.  Unfortunately, like for any wireless technology, disruption through radio jamming is possible.  This paper investigates the extent to which 5G NR is vulnerable to jamming and spoofing, by analyzing the physical downlink and uplink control channels and signals.  We identify the weakest links in the 5G NR frame, and propose mitigation strategies that should be taken into account during implementation of 5G NR chipsets and base stations.
\end{abstract}

\footnotetext[1]{This research was conducted primarily while the author was a doctoral student at Virginia Tech.}
\blfootnote{This is the author's version of the work. For citation purposes, the definitive version of record of this work is: M. Lichtman, R. M. Rao, V. Marojevic, J. H. Reed, and R. P. Jover, ``5G NR Jamming, Spoofing, and Sniffing: Threat Assessment and Mitigation,''IEEE International Conference on Communications (ICC) Workshops - 1st Workshop on 5G Wireless Security (5G-Security), pp. 1--6, May 2018.}
\section{Introduction}


Recently, the Third Generation Partnership Project (3GPP) released the specifications for 5G New Radio (NR), which is expected to be the primary 5G standard moving forward.  In addition to providing commercial communications services, cellular networks are used to broadcast emergency information, announcing natural disasters and other crises.  As we have seen with LTE, despite being designed for commercial communications, the latest cellular technology is often utilized for mission-critical applications such as public safety and military communications.  Just as we have become dependent on LTE, over the next decade we will likely become dependent on 5G NR, which is why we must ensure it is secure and available when and where it is needed.  Unfortunately, like any wireless technology, disruption through deliberate radio frequency (RF) interference, or jamming, is possible.

The objectives of this article are to outline and motivate the need for 5G network security and reliability research, by providing insights into physical (PHY) layer vulnerabilities of 5G NR and surveying mitigation techniques that can harden the PHY layer of next generation 5G deployments.  We individually analyze each physical control channel and signal, then compare them in order to identify the weakest link.  This paper concludes with a survey of mitigation techniques that should be taken into account during implementation, in order to mitigate the jamming attacks discussed in this paper.  

\section{Background of 5G NR}
\label{sec:background}

The 5G NR architecture is composed of components of LTE combined with a new radio access technology that is not backwards compatible with LTE.  5G NR is operable from  below 1 GHz to 100 GHz, making it the first generation of cellular technology with such a flexible frequency range.

As with LTE, the downlink (DL) uses standard orthogonal frequency-division multiplexing (OFDM) with a cyclic prefix.  The uplink (UL), however, has the option between normal OFDM just like the downlink, or DFT spread OFDM (DFT-s-OFDM), which is essentially the same that LTE uses in its uplink.  The DFT-s-OFDM mode does not support multi-stream transmissions and is intended for coverage-limited cases \cite{study_on_phy_layer_aspects}. 

\begin{figure}
	\centering
	\includegraphics[width=3.2in]{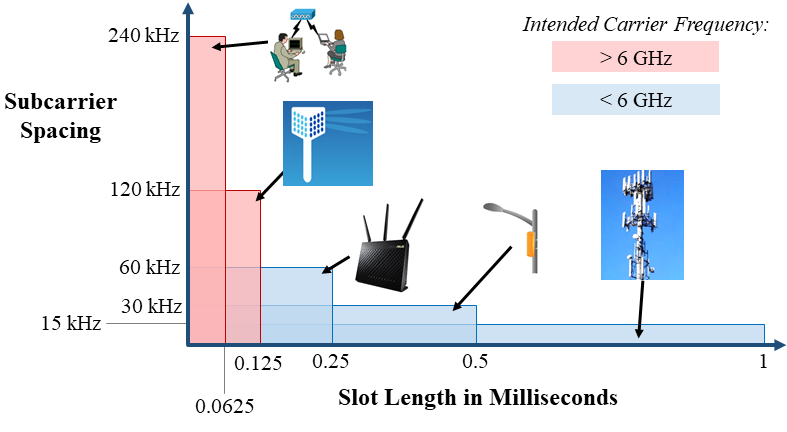}
	\caption{Unlike LTE, 5G NR defines a flexible frame structure and numerology.}
	\label{fig:numerologies}
\end{figure}

\begin{figure*}
	\centering
	\includegraphics[width=6.7in]{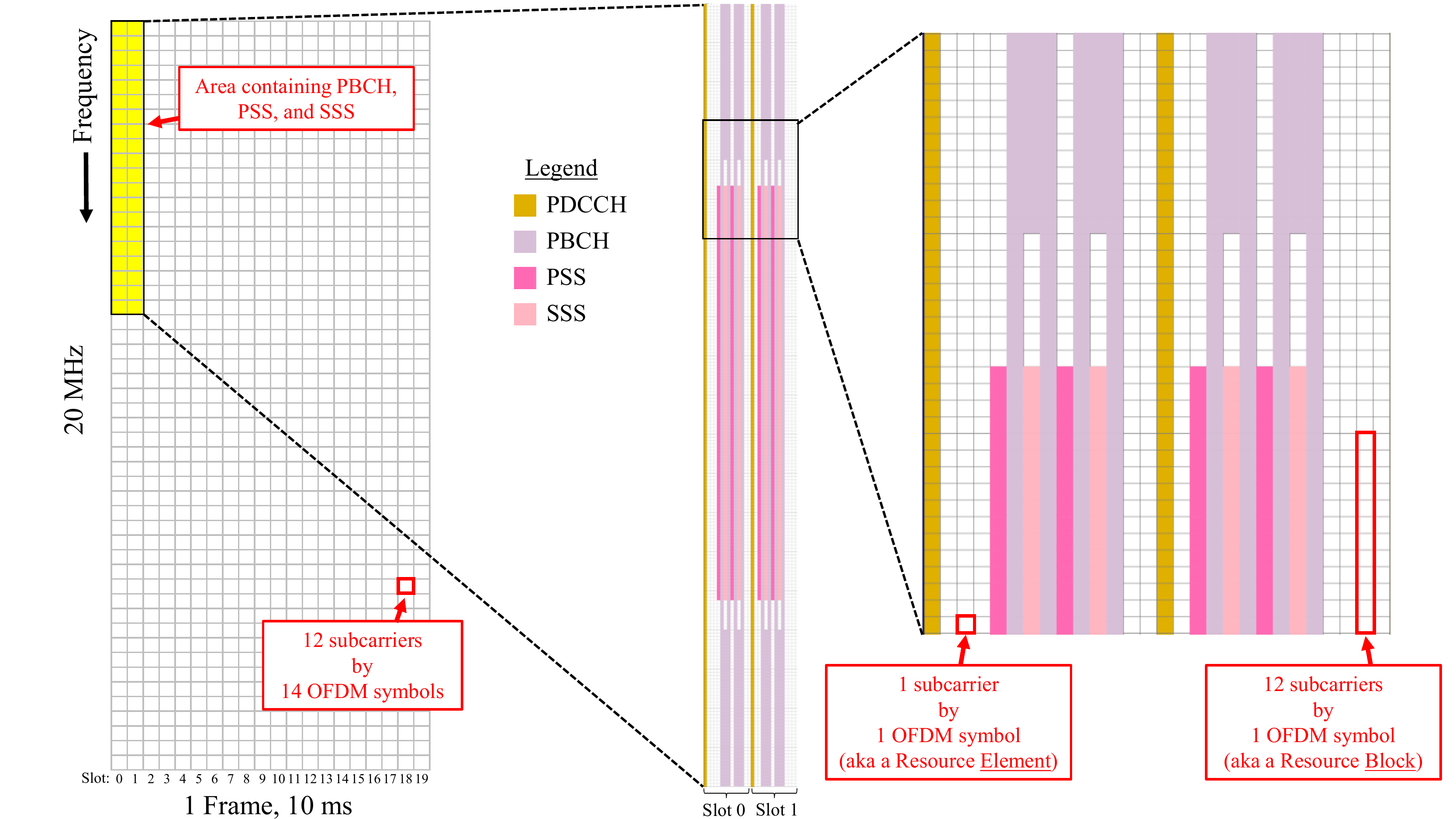}
	\caption{An example of the 5G NR downlink signal frame structure for a 20 MHz signal with 30 kHz subcarrier spacing, using a carrier below 3 GHz.}
	\label{fig:downlink}
\end{figure*}

5G NR supports frequency division duplexing (FDD) and time division duplexing (TDD).  
Figure \ref{fig:downlink} shows an example of the downlink frame. The 5G NR frame structure is similar to LTE, but incorporates much more flexibility and includes some important modifications.  Similar to LTE, every frame is 10 ms long in duration, there are 10 subframes in one frame, and there are 14 OFDM symbols in a slot. A major difference is that the number of slots per subframe (which was always equal to two in LTE), is now variable.  More specifically, the number of slots per frame is a function of the subcarrier spacing, which varies between 15 and 240 kHz. In Table \ref{bw-options} we list the different options and the carrier frequencies they are designed for.  

Similar to LTE, a Resource Element (RE) is one subcarrier by one OFDM symbol. Unlike LTE, the  Resource Block (RB) is 12 subcarriers by 1 OFDM symbol. Table \ref{rbs_per_bw} indicates the number of RBs as a function of system bandwidth and subcarrier spacing when below 6 GHz. 

\begin{table}[]
	\centering
	\small
	\renewcommand{\arraystretch}{1.2}
	\caption{Options in 5G NR for subcarrier spacing \cite{phy_channels_modulation}}
	\label{bw-options}
	\begin{tabular}{ccccc}
		\hline
		\begin{tabular}[c]{@{}c@{}}Subcarrier\\ Spacing\end{tabular} & \begin{tabular}[c]{@{}c@{}}Slots per\\ Subframe\end{tabular} & \begin{tabular}[c]{@{}c@{}}Meant for\\ Carriers...\end{tabular} & \begin{tabular}[c]{@{}c@{}}Min BW\\ {[}MHz{]}\end{tabular} & \begin{tabular}[c]{@{}c@{}}Max BW\\ {[}MHz{]}\end{tabular} \\ \hline
		15 kHz                                                       & 1                                                         & \multirow{3}{*}{\textless \ 6 GHz}                                & 4.32                                                       & 49.5                                                       \\
		30 kHz                                                       & 2                                                         &                                                                 & 8.64                                                       & 99                                                         \\
		60 kHz                                                       & 4                                                         &                                                                 & 17.28                                                      & 198                                                        \\ \cline{3-3}
		120 kHz                                                      & 8                                                         & \multirow{2}{*}{\textgreater \ 24 GHz}                             & 34.56                                                      & 396                                                        \\
		240 kHz                                                      & 16                                                        &                                                                 & 69.12                                                      & 397.44                                                     \\ \hline
	\end{tabular}
\end{table}

\begin{table}
	\scriptsize 
	\setlength\tabcolsep{1.2pt} 
	\renewcommand{\arraystretch}{1.2}
	\centering
	\caption{Number of RBs for different BWs and subcarrier spacings}
	\label{rbs_per_bw}
	\begin{tabular}{c|
			>{\columncolor[HTML]{EFEFEF}}c c
			>{\columncolor[HTML]{EFEFEF}}c c
			>{\columncolor[HTML]{EFEFEF}}c c
			>{\columncolor[HTML]{EFEFEF}}c c
			>{\columncolor[HTML]{EFEFEF}}c c
			>{\columncolor[HTML]{EFEFEF}}c c
			>{\columncolor[HTML]{EFEFEF}}c }
		\textbf{\begin{tabular}[c]{@{}c@{}} \\ Spacing\end{tabular}} & \textbf{\begin{tabular}[c]{@{}c@{}}5\\   MHz\end{tabular}} & \textbf{\begin{tabular}[c]{@{}c@{}}10\\   MHz\end{tabular}} & \textbf{\begin{tabular}[c]{@{}c@{}}15\\   MHz\end{tabular}} & \textbf{\begin{tabular}[c]{@{}c@{}}20\\ MHz\end{tabular}} & \textbf{\begin{tabular}[c]{@{}c@{}}25\\ MHz\end{tabular}} & \multicolumn{1}{c}{\textbf{\begin{tabular}[c]{@{}c@{}}30\\ MHz\end{tabular}}} & \textbf{\begin{tabular}[c]{@{}c@{}}40\\ MHz\end{tabular}} & \textbf{\begin{tabular}[c]{@{}c@{}}50\\ MHz\end{tabular}} & \textbf{\begin{tabular}[c]{@{}c@{}}60\\ MHz\end{tabular}} & \textbf{\begin{tabular}[c]{@{}c@{}}70\\   MHz\end{tabular}} & \textbf{\begin{tabular}[c]{@{}c@{}}80\\ MHz\end{tabular}} & \textbf{\begin{tabular}[c]{@{}c@{}}90\\   MHz\end{tabular}} & \textbf{\begin{tabular}[c]{@{}c@{}}100\\ MHz\end{tabular}} \\ \hline
		15 kHz                                                                & 25                                                         & 52                                                          & 79                                                          & 106                                                       & 133                                                       & 160                                                                           & 216                                                       & 270                                                       &                                                           &                                                             &                                                           &                                                             &                                                            \\
		30 kHz                                                                & 11                                                         & 24                                                          & 38                                                          & 51                                                        & 65                                                        & 78                                                                            & 106                                                       & 133                                                       & 162                                                       & 189                                                         & 217                                                       & 245                                                         & 273                                                        \\
		60 kHz                                                                &                                                            & 11                                                          & 18                                                          & 24                                                        & 31                                                        & 38                                                                            & 51                                                        & 65                                                        & 79                                                        & 93                                                          & 107                                                       & 121                                                         & 135                                                       
	\end{tabular}
\end{table}

\section{Physical Layer Vulnerabilities of 5G NR}

This section analyzes each specific physical channel and signal of 5G NR, according to 3GPP Release 15, to determine how vulnerable it is to jamming and spoofing.  The extent to which a physical channel or signal is vulnerable to jamming is highly influenced by the sparsity of that channel/signal with respect to the entire time-frequency resource grid \cite{lichtman2016lte}. A factor that reduces the vulnerability of a channel/signal is whether it is mapped to the time-frequency resource grid using a dynamic scheme that involves higher layer parameters (that a jammer may not know).  

5G NR has been designed with the option to be used above 24 GHz. Operating in the millimeter wave band alone improves the resilience to jamming. The development involved in building a jammer that targets cells above 24 GHz will take significantly more hardware knowledge, time, and money. Therefore, we keep the analysis in this paper specific to cells operating in sub-6 GHz bands.  

\subsection{Synchronization Signals}


Similar to LTE, 5G NR contains a Primary Synchronization Signal (PSS) and Secondary Synchronization Signal (SSS) which together are used for frame/slot/symbol timing as well as conveying the Physical Cell ID.  There are 1008 unique Physical Cell IDs in 5G NR, the PSS has three possible combinations and the SSS has 336 combinations. 

The PSS is made up of an m-sequence of length 127, mapped to a contiguous set of 127 subcarriers within the same OFDM symbol.  An m-sequence is a type of pseudorandom binary sequence that is spectrally flat except the DC term.  

The SSS is also a sequence of length 127, mapped to the same subcarriers as the PSS but a different OFDM symbol.  The SSS uses a Gold sequence, which is formed by combining two m-sequences.  Gold sequences within the same set have low cross-correlation, allowing a UE to distinguish between several nearby base stations on the same carrier  at low signal-to-noise-plus-interference ratio (SINR), making them resilient to jamming.  

Unlike LTE, the PSS and SSS are not always mapped to the downlink resource grid in the same location. The mapping depends on the cell's subcarrier spacing, the carrier frequency, and the parameter \textit{offset-ref-low-scs-ref-PRB} \cite{phy_channels_modulation}.  
There are three different options for subcarrier spacing below 6 GHz (15, 30, 60 kHz), and for all three options the PSS and SSS are mapped to the first two slots for carriers below 3 GHz, and the first four slots for those above 3 GHz. An example of PSS and SSS mapping is shown in Figure \ref{fig:downlink} for a carrier below 3 GHz.  

A jammer designed to jam the PSS and/or SSS selectively in time has to synchronize to the cell in time, and identify the subcarrier spacing (which might already be known beforehand using publicly available band plans).  This is only a little more complicated than PSS/SSS jamming in LTE.  

However, just as in LTE, it may be more effective for an adversary to transmit fake PSS/SSS signals rather than attempt to inject noise on top of the existing PSS/SSS, because it does not have to synchronize to a cell \cite{lichtman2016lte}.  It also uses less power because the PSS and SSS are designed to be detected at low signal-to-noise ratio (SNR), thus requiring more jammer power to successfully jam the signal.  Although it depends on the chipset and whether any PSS/SSS blacklisting mechanism exists, it may only be necessary to spoof the PSS.  Spoofing the PSS (and SSS) involves the attacker transmitting several fake PSS's, asynchronous to the target 5G NR frame(s) (i.e., not overlapping in time with the real PSS) and at higher power.  PSS/SSS spoofing can cause denial of service (DoS), which would likely occur during initial cell search \cite{marojevic2017performance}.

Similar to LTE, the 5G NR specifications do not specify the behavior of the UE when it detects a valid PSS with no associated SSS \cite{lichtman2013vulnerability}. Hence, the effects of PSS spoofing will be implementation-specific.  The more fake PSSs that are transmitted, the more sophisticated of a blacklisting mechanism is required for mitigation.  


\subsection{Physical Broadcast Channel}

The Physical Broadcast Channel (PBCH) is transmitted in the same slots as the PSS and SSS, an example of which is shown in Figure \ref{fig:downlink}.  The PBCH region spans more subcarriers than the PSS/SSS; it occupies 240 subcarriers, across 12 OFDM symbols for a carrier below 3 GHz, and 24 symbols for a carrier above 3 GHz. The information carried on the PBCH is known as the Master Information Block (MIB), which includes parameters such as the subcarrier spacing, position of downlink reference signals, and the position of downlink control channel.  All of this information is vital to a UE in order to attach to a cell.

\subsubsection{Jamming Vulnerability of the PBCH}
The symbols assigned to the PBCH region are all within two or four slots of each other (depending on whether the carrier is below or above 3~GHz respectively), so a jammer selectively targeting the PBCH will appear to have a very low duty cycle, especially at the higher subcarrier spacings where the duration of one slot is lower.  

By jamming the PBCH, UEs will not be able to access critical information they need to connect to a cell, thus preventing new UEs from accessing one or more cells.  
PBCH jamming can be performed in a time-selective manner if the jammer can synchronize to the target cell.  Otherwise, the jammer could simply jam the subcarriers the PBCH is on using 100\% duty cycle.  This latter approach involves jamming 240 subcarriers, and to provide some perspective, a 20 MHz downlink using 15~kHz subcarrier spacing has 1272 subcarriers.  Thus, it would involve jamming 19\% of the downlink signal, leading to a jamming gain around 7 dB w.r.t. barrage jamming.  

\subsubsection{Sniffing and Spoofing Vulnerability of the PBCH}
\label{sec:spoofing}

Despite the large number of changes in the PHY layer of 5G NR, most of the underlying protocol implementations are very similar to those of LTE starting in 3GPP Release 8. As a result, both the MIB and System Information Block (SIB) messages maintain a similar structure and payload to LTE. While the MIB message contains essential PHY layer configuration necessary by the UE to establish a radio link with a cell, the SIB messages contain detailed information on the configuration of the cell and overall network.  As described in \cite{5G_RRC}, the SIB messages provide information such as the idle timer configuration of the network, unique identifiers of the cell, and the RB mapping of critical control channels. Further details are included in the SIB messages which provide information on the received power thresholds that trigger a handover to another cell and similar mobility actions. This information, although not always necessary before the UE establishes a secure and encrypted connection with the network, is always broadcasted in the clear.  As previously discussed in the literature in the context of LTE, this information could be leveraged by an adversary.

It is worth noting that SIB and Radio Resource Control (RRC) messages in 5G NR include multiple new parameters, such as a list of cells to be added to a whitelist or blacklist. Considering that these messages occur prior to authentication and are not protected, it is likely that some of these fields could potentially be leveraged for security exploits against the 5G NR protocol. This could be achieved, for example, by spoofing SIB messages or impersonating a base station during the RRC handshake.

The 5G NR specifications also use a very similar RRC and Non-Access Stratum (NAS) protocol architecture as LTE. 
As a result, there is still a number of pre-authentication messages implicitly trusted by both the UE and the base station. By both impersonating a UE or a base station, an adversary can leverage exploits already known in LTE networks \cite{jover2016exploits,LTEpracticalattacks}. 

\subsection{Physical Downlink Control Channel}

The Physical Downlink Control Channel (PDCCH) is used to send control information to the UEs on a per-slot basis. It is used to schedule downlink transmissions, uplink transmissions, modulation and coding format of those transmissions, and hybrid-ARQ information \cite{layer_three_overall_description}. The PDCCH can appear on any subcarrier; so the jammer must decode the parameter \textit{CORESET-freq-dom} \cite{phy_channels_modulation}. The parameter \textit{CORESET-time-dur}, which can take on values 1, 2, or 3, indicates how many OFDM symbols the PDCCH occupies each slot \cite{phy_channels_modulation}. The PDCCH always starts in the first symbol of each slot, is QPSK modulated and uses polar coding. 

In order to jam all possible locations the PDCCH might reside in, assuming knowledge of \textit{CORESET-freq-dom}, the jammer would have to jam every subcarrier, using a duty cycle of either 7\%, 14\%, or 21\% depending on the value of \textit{CORESET-time-dur}.  This type of pulsed jamming attack can also act as a form of automatic gain control jamming \cite{adamy2001ew}.  

\subsection{Reference Signals}

5G NR includes several reference signals (RSs) which act as pilots.  Unlike LTE, Demodulation RS (DM-RS) are separated by physical channel, because 5G NR does not include cell specific reference signals.  There is a separate DM-RS for the PDSCH, PUSCH, PDCCH, PUCCH, and PBCH.  In addition, there is the Phase-tracking RS (PT-RS) for the PDSCH and PUSCH, Channel state information RS (CSI-RS) for the downlink, and Sounding RS (SRS) for the uplink \cite{phy_channels_modulation}.  We will focus on the RSs that are most vulnerable to jamming.


From the perspective of the jammer, the best RS to jam would be the one that requires the least amount of energy to jam (i.e. least number of REs per frame), but also be vital to the operation of the link.  The DM-RS for the PBCH fits the bill, because it is in the same spot every frame and only requires knowledge of the cell ID and where the PBCH is located, which can easily be known if the jammer is already able to time-synchronize to the frame.  The DM-RS for the PBCH occupies \sfrac{1}{4} of the REs assigned to the PBCH.  It is also possible to jam the DM-RS for the PBCH without needing to synchronize to the cell, by jamming the correct 60 subcarriers.  

Phase-tracking RS for the PDSCH are only used when the higher layer parameter \textit{Downlink-PTRS-Config} is set to ON, and even then the mapping is a function of parameters \textit{timeDensity} and \textit{frequencyDensity}.  The effectiveness of a downlink PT-RS jamming attack is not clear using the information currently available, because we would need to know how often PT-RS are enabled in practice, and what density the base station vendors decide to use as a default. 

\subsection{Downlink and Uplink User Data}

The Physical Downlink Shared Channel (PDSCH) and Physical Uplink Shared Channel (PUSCH) are used to transmit user data from the base station to the UE and vice versa, and represent the bulk of the frame.  While surgically jamming these channels is possible, the adversary might as well jam the entire uplink or downlink signal (and hence this is what we show in Figure \ref{2d_ranking}).  Thus, PDSCH and PUSCH jamming are two of the least important threats to consider.


\subsection{Physical Uplink Control Channel}

The Physical Uplink Control Channel (PUCCH) is used by the UE to send the base station a variety of control information, including hybrid-ARQ acknowledgments, scheduling requests, and channel state information \cite{layer_three_overall_description}.  There are five different PUCCH formats, and there are a variety of parameters provided by the higher layer to inform the UE about which subcarriers and symbols to transmit each PUCCH message on.  
The PUCCH has an option for intra-slot hopping, which acts as a great defense mechanism.  The PUCCH is modulated with either BPSK or QPSK and uses either repetition code, simplex code, Reed Muller code, or polar code (depending on the number of bits to transmit).  Similar to LTE, uplink control information can also be carried on the PUSCH, meaning jamming just the PUCCH will not block all uplink control information.  All of these factors make the PUCCH an extremely complicated physical channel to jam.

\subsection{Physical Random Access Channel}

The 5G NR random access procedure for regular UEs is very similar to that of LTE.  When a UE wants to connect to a cell, it first receives the PSS, SSS, and PBCH.  After synchronizing to the cell in time and frequency, it transmits a preamble over the Physical Random Access Channel (PRACH), which takes the form of a Zadoff-Chu sequence that embeds a value used to temporarily identify the UE.  The base station broadcasts the candidate locations of the PRACH (in time and frequency) in case a UE wants to connect.  The possible configurations are listed in Table 6.3.3.2-2 of \cite{phy_channels_modulation}.  As with LTE, the large number of possible locations, and the complexity required to decode the positions in real time, makes the PRACH an unlikely target for a jammer. On the other hand PRACH spoofing, where the UEs flood the PRACH might be feasible, but the 5G NR specifications do not specify the behavior of the base station when encountering a large number of invalid preambles.

\section{Vulnerability Assessment}

This section compares the attacks in terms of power efficiency and complexity to quantify the vulnerability of 5G NR, as a basis for developing effective protocol improvement and hardening techniques. First, we need to introduce two different ways of measuring the received jammer-to-signal ratio (J/S), that is, the ratio of the received jamming signal power to the received 5G NR signal power.  

Following the convention in \cite{lichtman2016lte}, we define ${J/S}_{CH}$ as the J/S that only takes into account the specific REs of the channel or signal being jammed. For example, when jamming the broadcast channel (the light purple region in Figure \ref{fig:downlink}), it is assumed the jammer will place its energy on top of the broadcast channel in time and frequency, and not transmit on any other REs (this is an ideal case from the perspective of the jammer). Thus, ${J/S}_{CH}$ in this instance corresponds to the average received signal power from the jammer divided by the averaged received power of only the broadcast channel.

J/S averaged over an entire radio frame is referred to as ${J/S}_F$. Using the previous example of jamming the broadcast channel, ${J/S}_F$ corresponds to the received power from the jammer accumulated over one frame, divided by the received power of the accumulated signal power over the entire 10 ms UL or DL frame. Our calculations of ${J/S}_F$ assume FDD, since the TDD mode has a lot of configurations and hence, requires a separate analysis. The ${J/S}_F$ metric provides a convenient way to compare each jamming attack against the baseline attack, which is jamming the entire DL or UL frame. 

Note that J/S alone does not provide enough information to determine the affected area. Link budgets, which take into account factors like the jammer's transmit power and channel attenuation, are needed to determine such information.

The vulnerability of each channel or signal is based primarily on three factors:
\begin{enumerate}
	\item The sparsity of the channel/signal w.r.t. the entire downlink or uplink frame, i.e. the percent of REs occupied by the channel/signal.
	\item The jamming power needed to significantly corrupt the channel or signal.
	\item The complexity of the jammer required to perform such an attack, based on whether synchronization to the cell is needed and whether parameters need to be decoded.
\end{enumerate}

This information for each channel and signal is summarized in Table \ref{phy_channels}. The sparsity can be combined with the minimum ${J/S}_{CH}$ needed to estimate the corresponding ${J/S}_{F}$. This is an approximation because it assumes a uniform power spectral density across the 5G NR downlink or uplink signal. From the perspective of a jammer trying to minimize its power consumption, a lower ${J/S}_{F}$ is better.  To calculate the ``\% of REs'' column, which then gets used to calculate ${J/S}_{F}$, we have assumed a carrier frequency below 3 GHz, a channel bandwidth of 20 MHz, and a subcarrier spacing of 30 kHz.  

\begin{figure}
	\centering
	\includegraphics[width=3.3in]{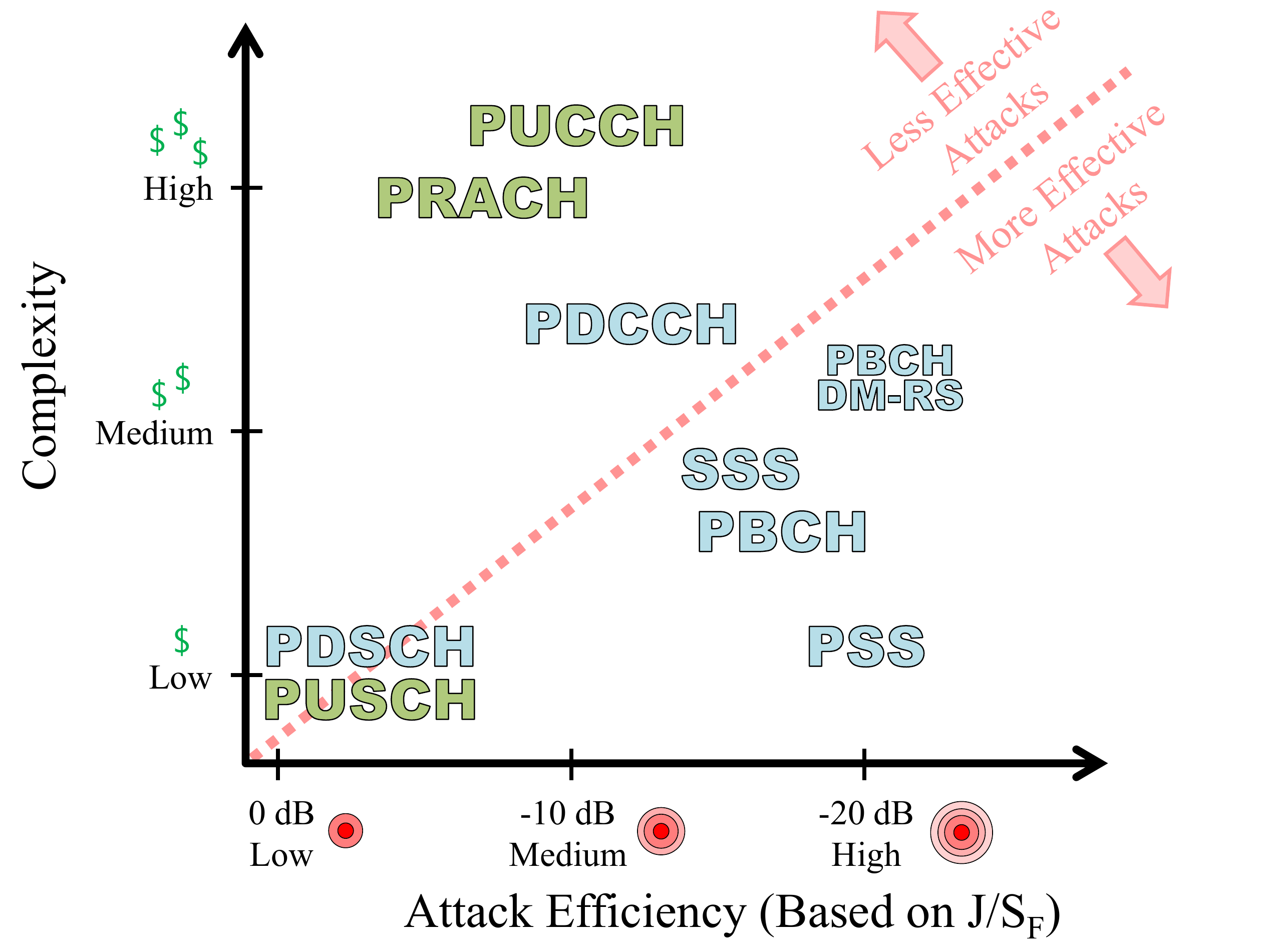}
	\caption{Ranking of 5G PHY-layer attacks based on efficiency and complexity.}
	\label{2d_ranking}
\end{figure}

\begin{table*}
	\centering
	\caption{Channel and signal modulation/coding, sparsity, synchronization and cell params requirement, and minimum J/S to cause DoS}
	\label{phy_channels}
	\rowcolors{2}{gray!25}{white}
	\begin{tabular}{cccccccc}
		\rowcolor{gray!50}
		\textbf{Channel/Signal} & \textbf{Modulation} & \textbf{Coding} & \textbf{\% of REs} & \textbf{Synch. Required} & \textbf{Params. Required} & $\boldsymbol{{J/S}_{CH}}$ & $\boldsymbol{{J/S}_{F}}$ \\
		\hline
		PDSCH (Downlink) & \{4, 16, 64, 256\}-QAM& LDPC   		& 90\%				   & 	No	& None  & 0 dB  & -1 dB     \\
		PBCH 		 	 & QPSK 					& Polar			& 1.7\% 			&	Yes	& None  & 0 dB  & -17 dB 	\\
		PDCCH 		 	 & QPSK 					& Polar			& 7\% 				&	Yes	& Medium  & 0 dB  & -11 dB 	\\
		PUSCH (Uplink) 	 & \{4, 16, 64, 256\}-QAM& LDPC  & $\sim$ 90\% 		&	No	            & None  & 0 dB  & -1 dB 	\\
		PUCCH 			 & QPSK 		 	        & Variety       & $\sim$ 10\%   	&	Yes	& High  & 0 dB  & -10 dB 	\\
		PRACH 			 & Zadoff–-Chu Sequence 	& N/A 			& $\sim$ 2\% 		&	Yes	& Medium  & 10 dB & -7 dB 	\\
		PSS (Spoofing) 	 & M-Sequences 			    & N/A 			& 0.1\% (3 PSSs)	&   No	& None  & 10 dB & -20 dB 	\\
		SSS 			 & Gold Sequences	 	    & N/A 			& 0.3\%  			&	Yes	& None  & 10 dB & -15 dB 	\\
		PBCH DM-RS   	 & QPSK 		 			& N/A 			& 0.4\%  		    &	Yes	& Low  & 3 dB  & -21 dB 	\\
	\end{tabular}
\end{table*}

In order to approximate ${J/S}_{CH}$ needed for each attack to cause DoS, we must find the SINR at which the modulation and coding scheme for that channel causes a significantly high error rate.  Most of the control channels discussed use polar codes, at various rates.  It can be shown that around 0 dB SNR, polar codes at rate \sfrac{1}{3} reach an extremely high bit error rate \cite{ercan2017error}.  For the synchronization signals, it must be corrupted to the point of causing an extremely low probability of detection.  Because of the number of factors involved, these are only approximations, although we will show that our ${J/S}_{CH}$ values only make up a small portion of the ${J/S}_{F}$ metric; channel sparsity is typically the decisive factor.  

Based on the information provided in Table \ref{phy_channels}, we can form an initial threat assessment of the vulnerability of 5G NR to jamming and RF spoofing. 
Because there is an efficiency and complexity aspect to each attack, instead of simply ranking them, we have assembled the attacks into a two-dimensional plot that is shown in Figure \ref{2d_ranking}. From the perspective of a jammer, the most effective attacks are towards the bottom-right. Specifically, we believe that efforts toward hardening 5G NR for critical communications should focus on addressing possible PSS spoofing and PBCH jamming attacks.  Compared to the same analysis of LTE (see \cite{lichtman2016lte}, \cite{Rao_MILCOM_2017}, \cite{lichtman2016communications}), 5G NR contains fewer ``extremely vulnerable'' components.  Specifically, LTE's PCFICH and PUCCH are especially vulnerable to jamming. The PCFICH has been removed for 5G NR, and the PUCCH is now mapped in a more dynamic manner, so that it cannot be jammed simply by transmitting energy on the outer edges of the uplink band.

It is important to understand that even the more complex attacks can be implemented with widely available open-source libraries, a low-cost SDR with a budget under \$1000, and basic Linux programming skills \cite{Rao_MILCOM_2017}. It is for this reason that jamming mitigation should be strongly considered.

\section{Survey of Mitigation Techniques}
\label{sec:mitigation}

In this section we briefly describe some mitigation techniques that do not require changes to the specifications, but instead can be incorporated into implementations of 5G NR chipsets and base stations. The technology inside of modern cellphones and other UEs resides in the chipset. On the other side of the link, the base stations typically use a baseband unit that does most of the processing in software, and an RF module handles the RF processing chain. Thus, changes to the behavior of the base stations likely only require a software update, whereas changes to the UE require a new chipset to be designed and manufactured.

PSS spoofing can be mitigated by creating a timer for receiving the SSS. If this timer expires, the UE should blacklist the PSS for a certain amount of time, and choose the second strongest cell within the same frequency. PSS and SSS spoofing attacks can be mitigated by having the UE create a list of all available cells in the given frequency channel along with their received power levels \cite{Mina_CellSelect_2017}. The UE could then search for the PBCH of the strongest cell and have another timer for decoding the MIB. If this timer expires, the UE would look for the PBCH of the next strongest cell, and so forth.  One important requirement for this approach is that the blacklist not be limited to a certain number of ``bad'' cells, and that it has some sort of knowledge decay so that an adversary transmitting a different PSS every frame (both in terms of the sequence and the timing offset) does not saturate the abilities of the blacklisting mechanism. 

Regarding sniffing and spoofing potential threats, 5G NR still lacks the necessary protection to security threats that were evident in LTE networks \cite{jover5Gfilling}. Broadcast messages, particularly MIB and SIB packets, contain a myriad of information, not all of which is necessary for the UE to establish a secured connection. Ideally, SIB message content would be limited to strictly what is necessary to establish a radio link with the base station, and further network configuration elements would be shared on a secured and integrity protected broadcast channel. Moreover, both UEs and base station implicitly trust all messages prior to authentication and encryption establishment, which could lead to well know security exploits. It is necessary to propose methods that allow a UE to determine whether a base station is legitimate prior to executing certain procedures based on the unauthenticated RRC and NAS messages, even though the specifications do not require such mechanism.  

These mitigation strategies only address a few of the attacks that we discussed in this article. Further research on mitigation techniques is needed.  Developing solutions early on will ensure a cost-effective and timely deployment and penetration into many markets, as opposed to dealing with the consequences of potential network failures in the future. Innovative solutions are sought that lead to conceptual changes to how mission-critical 5G networks and UEs of the future are deployed and operated.

\section{Conclusion}

In this article we analyzed the vulnerability of 5G NR to jamming, spoofing, and sniffing by looking at individual physical channels and signals. Using barrage jamming as a baseline, we have shown that more effective jamming methods can be realized by exploiting the protocol. We used well-established metrics related to the efficiency and complexity of each method to compare them and conclude that the PSS and PBCH are the weakest subsystems and should, therefore, be addressed first. Compared to LTE, 5G NR is far less vulnerable to jamming, mainly because of its dynamic nature and removal of sparse control channels like the PCFICH. If 5G NR follows the same pattern as 4G LTE, we can expect it to be highly relied-upon during the next decade.  We therefore recommend that the identified mitigation techniques be considered before deployment. Lastly, additional research is needed for testing and advancing the 5G NR standard in terms of physical layer security.

\section*{Acknowledgments}
The work of Rao, Marojevic, and Reed was funded in part by the National Science Foundation (NSF) under Grant CNS-1642873, and the Oak Ridge National Laboratory. 
\bibliographystyle{IEEEtran}
\bibliography{references}

\end{document}